\documentclass[twocolumn,showpacs,preprintnumbers,amsmath,amssymb,epsf]{revtex4-1}

\usepackage{graphicx,epsfig}
\usepackage{dcolumn}
\usepackage{bm}
\newcommand{\be}{\begin{equation}}
\newcommand{\ee}{\end{equation}}
\newcommand{\bse}{\begin{subequations}}
\newcommand{\ese}{\end{subequations}}
\newcommand{\bary}{\begin{eqnarray}}
\newcommand{\eary}{\end{eqnarray}}
\newcommand{\bwt}{\begin{widetext}}
\newcommand{\ewt}{\end{widetext}}

\begin{document}


\title{Blazar origin  of some IceCube events}
\author{Luis Salvador Miranda, Alberto Rosales de Le\'{o}n, Sarira Sahu} 

\affiliation{
Instituto de Ciencias Nucleares, Universidad Nacional Aut\'onoma de M\'exico, 
Circuito Exterior, C.U., A. Postal 70-543, 04510 Mexico DF, Mexico\\
}

\begin{abstract}

Recently ANTARES collaboration presented a time dependent analysis 
to a selected number of flaring blazars to look for 
upward going muon events produced from the charge current interaction
of the muon neutrinos. We use the same list of flaring blazars to
look for possible positional correlation with the IceCube neutrino
events. In the context of photohadronic model we propose that the neutrinos are
produced within the nuclear region of the blazar where Fermi accelerated
high energy protons interact with the background synchrotron/SSC
photons. Although we found that some  objects from the ANTARES list are within the error
circles of few IceCube events, the statistical analysis shows that  none of these sources have a 
significant correlation.

\end{abstract}

\maketitle

\section{Introduction}

Interaction of ultra high energy cosmic rays (UHECRs) with the background medium
(photons and protons) produce high energy $\gamma$-rays and
neutrinos. On their way to Earth the UHECRs can be deflected in the magnetic field and the high energy
$\gamma$-rays can be absorbed. So both of these
heavenly messengers will lose their directionality. On the other hand
neutrinos will be directly pointing to the source, that is why
neutrinos are considered as ideal cosmic messengers.

The IceCube detector located at South Pole in Antarctic ice is
precisely built to look for high energy neutrinos (above few TeV) 
by measuring the Cherenkov radiation of the secondary particles
created in each neutrino event. The energy deposited by each event,
their direction and topology can be calculated from the
trail of the observed Cherenkov light. 
In 2012 the IceCube Collaboration published two years of data
(2010-2012) in which 28 neutrino events with energies between 30 to 1200
TeV were observed\cite{Aartsen:2013jdh}. Twenty one of these events are shower-like and the
rest are muon tracks. In this analysis two events were PeV neutrino shower
events. Adding a third year of analysis in a total 988-days data
revealed a total of 37 events, of which 9 are
track events and the rest are shower events\cite{Aartsen:2014gkd}. 
The shower events have large angular errors (an average of $15^{\circ}$) than
the track events (about $1^{\circ}$). These events have flavors,
directions and energies inconsistent with those expected from the
atmospheric muon and neutrino backgrounds. So the study of arrival
directions are helpful to find sources of high energy neutrinos
and the relevant acceleration mechanism acting within the source.

The isotropic distribution of these IceCube neutrino events suggest
contribution from at least
some extragalactic sources. There exist different types of potential
astrophysical sources to produce UHECRs and hence high energy
neutrinos and $\gamma$-rays. The list includes:
Gamma-ray bursts (GRBs)\cite{Murase:2013ffa}, core of active
galactic nuclei (AGN)\cite{Winter:2013cla}, high energy peaked blazars
(HBLs)\cite{ Padovani:2014bha, Krauss:2014tna,Sahu:2014fua},
starburst galaxies\cite{Murase:2013rfa} and 
sources from Galactic center\cite{Razzaque:2013uoa}.
In Ref.\cite{Padovani:2014bha} many positional correlations of BL Lac
objects and galactic pulsar wind nebulae with the IceCube events are
shown. There are also nonstandard physics interpretations
of these IceCube events from the decay of superheavy dark matter
particles, leptoquark interaction and decay of exotic neutrinos\cite{Chen:2013dza}
(See \cite{Anchordoqui:2013dnh}  for a recent review).

Recently ANTARES collaboration presented a time dependent analysis\cite{Adrian-Martinez:2015wis} to
look for upward going muon tracks by charge current interaction of 
$\nu_{\mu}$  from flaring blazars selected
from the Fermi-LAT and TeV $\gamma$-ray observed by ground based
telescopes H.E.S.S, MAGIC and
VERITAS respectively. In this analysis the most significant correlation
was found with a GeV flaring blazar from the Fermi-LAT
catalog. However, the post-trial probability
estimate shows that the event was compatible with background
fluctuations. 
In this work we would like to analyse the above list of Fermi-LAT flaring blazars to see
if there is any correlation with the IceCube
neutrino events. We use the unbinned maximum likelihood method (MLM)
with two different values of the spectral index 
for our analysis of the positional correlation of these objects. 

\section{Candidates}

Blazars are believed to be the most likely candidates to produce
UHECRs and neutrinos\cite{ Padovani:2014bha,
  Krauss:2014tna,Sahu:2014fua}. 
These are extragalactic objects characterised by relativistic jets
with a small viewing angle with respect to the line of sight and are
powered by a supermassive black hole in the center of their respective
galaxy. These objects are also 
efficient accelerators of
particles through shock or diffusive Fermi acceleration processes
with a power-law spectrum given as ${dN}/{dE} \propto E^{-\kappa}$,
with the power index $\kappa \geq 2$\cite{Dermer:1993cz}.
Protons can
reach ultra high energy through the above acceleration
mechanisms. Fractions of these particles escaping from the source
can constitute the UHECRs arriving on Earth.
These objects also produce  high 
energy $\gamma$-rays and neutrinos through $pp$ and/or $p\gamma$ interactions 
\cite{Kachelriess:2008qx}.
The classification for these sources are according to the properties of
their emission lines: if a strong
broad emission line in the optical spectrum is present, it is classified as Flat
Spectrum Radio Quasar (FSRQ), otherwise is a BL Lacerate (BL Lac)
object. Depending on the frequency of the first peak, the BL Lac objects are
further classified into low (LBL), intermediate (IBL) and high energy
(HBL) peaked objects. 

The ANTARES collaboration searched
for high energy cosmic muon neutrinos using the data taken during the
period August 2008 to December 2012. The collaboration selected 
41 very bright and variable Fermi-LAT blazars with significant time
variability and having the  flux 
$>\, 10^{-9}$ photons ${cm^{-2}\,s^{-1}}$ for  the $\gamma$-ray energy
above 1 GeV. They have also selected seven TeV flaring objects reported by
H.E.S.S., MAGIC and VERITAS telescopes with the expectation that the
TeV $\gamma$-rays may be correlated with the neutrino events. From the
41 Fermi blazar list,  33 are FSRQs, 7 are BL Lacs and one is
unknown. Similarly from the list of 7 TeV flaring blazars one is FSRQ and six
are HBLs. It shows that both FSRQs and HBLs are
probable sources of very high energy neutrinos and can be possible
sources for some of the IceCube event. 
It is suggested that UHECRs are accelerated in the inner jet of
FSRQ and interact with the background from the broad-line region
(BLR), synchrotron radiation or the photon 
from accretion disk\cite{Murase:2014foa,Dermer:2014vaa,Wang:2015woa}.

In a previous article\cite{Sahu:2014fua} we proposed that photohadronic interactions of
the Fermi accelerated high energy protons with the 
background photons in the
nuclear region of the HBLs and AGN are responsible for
some of the IceCube events. These objects were observed in multi-TeV
$\gamma$-rays  and some had also flaring.  
In this model it is assumed that the flaring of blazar in high energy $\gamma$-ray
occurs within a compact and confined
region with a comoving radius $R'_f$ inside the blob of radius
$R'_b$\cite{Sahu:2013ixa} (henceforth $'$ implies jet comoving frame). In the inner region, the photon
density $n'_{\gamma,f}$ is very high compared to the photon density $n'_{\gamma}$ in
the outer region i.e. $n'_{\gamma,f} \gg n'_{\gamma}$. 
Fermi accelerated
high energy protons undergo photohadronic interaction with the
seed photons in the inner region in the self-synchrotron Compton (SSC) regime through the
intermediate $\Delta$-resonance. 
On the other hand, in a normal blazar jet, the
photohadronic process is not an efficient mechanism to produce
multi-TeV  $\gamma$-rays and neutrinos because $n'_{\gamma}$ is low, 
which makes the optical depth $\tau_{p\gamma}\ll 1$. But the
assumption of 
compact inner jet region overcome this problem where 
the optical depth of
the $\Delta$-resonance process is 
$\tau_{p\gamma}=n'_{\gamma,f} \sigma_{\Delta} R'_f$ and 
$n'_{\gamma,f} $ is unknown. 
We can estimate the photon density in this region 
by assuming that the Eddington luminosity is equally shared by the jet
and the counter jet in the blazar.
For a given comoving photon energy
$\epsilon'_{\gamma}$ in the synchrotron/SSC regime we can get the
upper limit on the photon density as  
$n'_{\gamma,f} \ll L_{Edd}/(8\pi R'^2_f \epsilon'_{\gamma})$. Also 
by comparing the proton energy loss time scale
$t'_{p\gamma}\simeq (0.5\,n'_{\gamma,f}\sigma_{\Delta})^{-1}$ and the
dynamical time scale  $t'_{d}=R'_f$ we can estimate
$n'_{\gamma,f}$, so that 
the production of multi-TeV $\gamma$-rays and
neutrinos take place. 
Not to have over production of neutrinos and $\gamma$-rays, we
can assume a moderate efficiency (a few percents)  by taking 
$\tau_{p\gamma} < 1$ which gives $n'_{\gamma,f} < (\sigma_{\Delta}
R'_f)^{-1}$. In this work we assume $1\%$ energy loss of the UHE
protons in the inner region on the dynamical time scale $t'_d$
corresponding to a optical depth of $\tau_{p\gamma} \sim 0.01$ and 
$n'_{\gamma,f}\sim 2\times 10^{10}\, R'^{-1}_{f,15}\, cm^{-3}$. Here
the inner blob radius $R'_f$ is expressed as $R'_f=10^{15} R'_{f,15}\,
cm$ and $R'_{f,15}\sim 1$\cite{Sahu:2014fua}.

In the photohadronic interaction, the intermediate
$\Delta$- resonance produced will give both high energy neutrinos and
$\gamma$-rays and relation between the seed photon and the neutrino
energy is given by 
\be
E_{\nu}\epsilon_\gamma =0.016 \frac{\Gamma \delta}{(1+z)^2} GeV^2,
\ee
where $E_{\nu}$ and $\epsilon_{\gamma}$ are respectively the observed
neutrino energy and the background photon energy. The source is
located at a redshift  $z$ and the bulk Lorentz factor of the jet is 
$\Gamma$. The Doppler factor is given by $\delta$. But for FSRQ and
BL Lac objects $\Gamma\simeq \delta$. So if $z$ and $\Gamma$ of a blazar
are known we can estimate the $\epsilon_{\gamma}$ from the given $E_{\nu}$.
The neutrino flux is given as\cite{Moharana:2015nxa}
\be
F_{\nu}=\sum_{\alpha} \int_{E_{\nu 1}(1+z)}^{E_{\nu 2}(1+z)}dE_{\nu}
E_{\nu} J_{\nu_{\alpha}} (E_{\nu}),
\label{fluxnu}
\ee
where for all neutrino flavors $\alpha$ ($e$, $\mu$ and $\tau$), a power-law spectrum of the
form
\be
J_{\nu_{\alpha}}(E_{\nu}) = A_{\nu_{\alpha} }\left (\frac{E_{\nu}} {100\, TeV}\right )^{-\kappa}
\ee
is taken. The normalization constant $A_{\nu_\alpha}$ is given by 
\be
A_{\nu_\alpha}=\frac{1}{3} \frac{N_{\nu}} 
{T  \Sigma_{\alpha} 
\int_{E_{\nu_1}}^{E_{\nu_2}} dE_{\nu} A_{eff,\alpha} (E_{\nu} )
\left ( 
\frac{E_{\nu}} {100 TeV}\right )^{-\kappa}
},
\ee
where $N_{\nu}$ is the number of neutrino events and $A_{eff,\alpha}$
is the effective area for different neutrino flavors. The energy
integrals are done in the limit 25 TeV to 2.2 PeV.
The time period
$T=988$ days is used\cite{Aartsen:2013jdh} for the calculation of normalization constant.

\section{Unbinned Maximum Likelihood Method}

To identify the possible sources of IceCube events we employ the
Unbinned Maximum Likelihood Method (MLM)\cite{James:2006zz} to find
spatial correlation between the blazar sample under consideration and
the IceCube events. The signal and the background weights are not
separable for an object and both contribute to the likelihood
function, which is given by the product of the individual probability
densities for the IceCube events as\cite{Braun:2008bg}
\be
 {\cal L} (n_{s}, \vec{x}_{s}) = \prod^{N}_{i=1} \left[\frac{n_s}{N} 
S_i (\vec{x}_{s}) + \left(1-\frac{n_s}{N}\right) B_i \right],
\ee
where N is the number of IceCube events we take into account, 
$n_s/N$ is the weight associate with the signal probability density function (PDF) and its values
vary between 0 and 1. The background PDF depends on the
the neutrino energy and the declination which is expressed as
\be
B_i = {\cal B}(E_i,\delta_i).
\ee
The background is constructed from the integrated effective areas of
the IceCube 79 strings configuration\cite{Aguilar:2013tc}. The
neutrino effective area depends on the detector 
geometry and the absorption of
the neutrinos by the Earth.  The background PDF takes into account the
contribution from the atmospheric muon neutrinos. Above $\sim 100$
TeV, neutrinos from  the decay of charm hadrons $D^{\pm}, D^0$ contribute to the
background neutrino flux known as prompt flux. Equal number of
neutrinos and anti-neutrinos of electron and muon flavors are produced
in this process. However, the prompt flux is poorly understood in the
high energy limit. For the background calculation we also include the
contribution from the prompt
background\cite{Enberg:2008jm,Desiati:2010wt}.

The signal PDF is defined as the  product of a spatial term and the
energy term as shown below
\be
S_i={\cal S}_i(|{\bf x}_i- {\bf x}_s|,\sigma_i)\, {\cal E}_i
(E_i,\delta_i,\kappa), 
\ee
where we have defined 
\be
{\cal S}_i (\vec{x}_{s})= \frac{1}{2 \pi \sigma^{2}_i} e^{- \frac{| x_i -
     x_s |^2}{2 \sigma_i^2} },
\label{spdf}
\ee
which is a Gaussian function\cite{Neunhoffer:2004ha}. In the above Eq.(\ref{spdf}), $| x_i -
x_s |^2$ is the space angle difference between the source and the
reconstructed event direction and $\sigma_i$ is the standard deviation
of the $i^{th}$ IceCube angular error distribution.
We also define
\be
\delta\chi^2=\frac{|x_i-x_s|^2}{\Delta},
\ee
The value of $\delta\chi^2 \le 1$ signifies that the object is inside the
median angular error $\Delta$ of the IceCube event.
The signal energy PDF ${\cal E}_i$ depends on
the event energy, spectral index $\kappa$ and the declination. Here we
use $\kappa=2$ and 2.5 for our analysis. 

\begin{figure}[t!]
{\centering
\resizebox*{0.56\textwidth}{0.40\textheight}
{\includegraphics{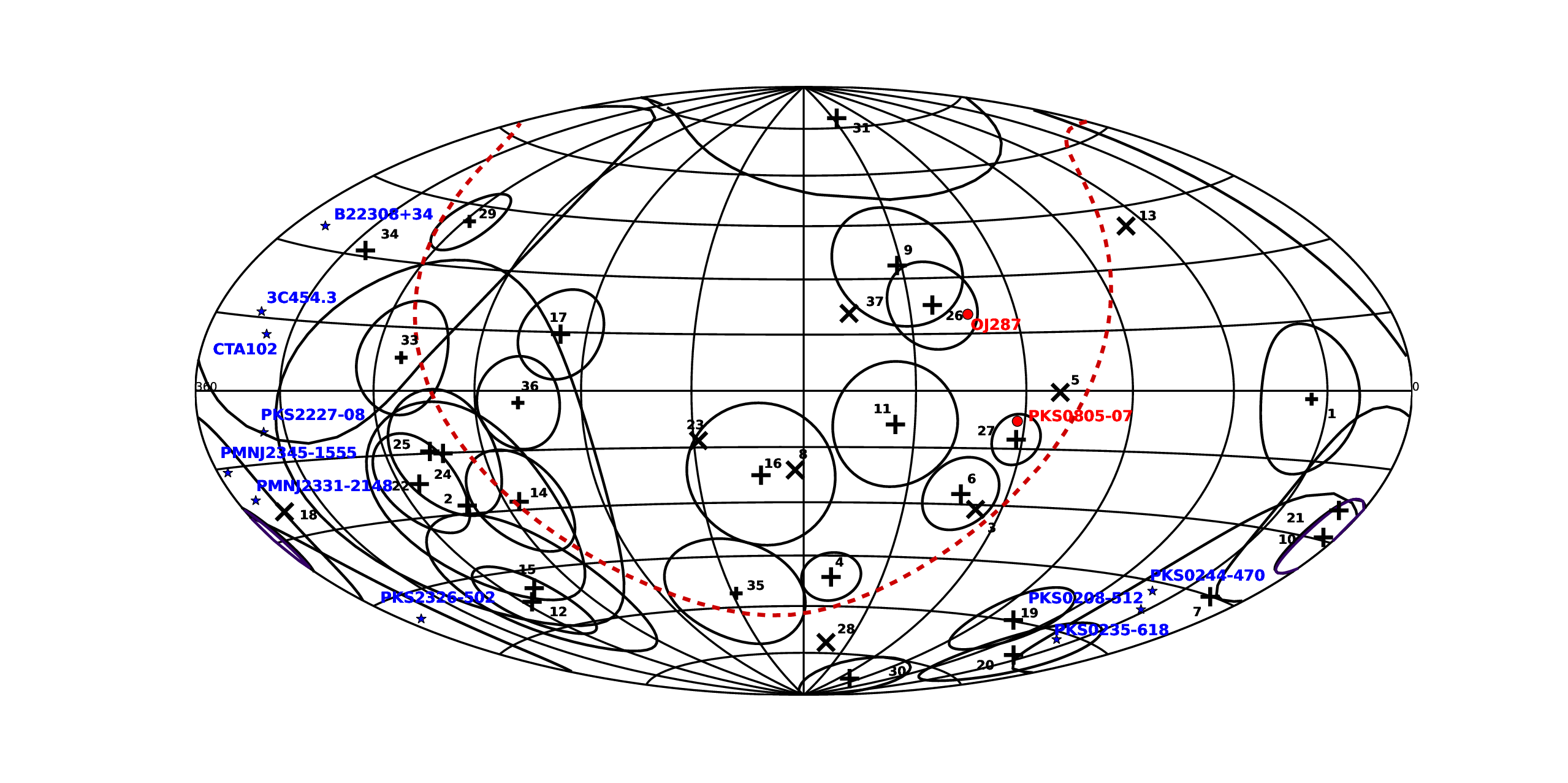}}
\par}
\caption{The sky map is shown in Equatorial coordinates with 37
  IceCube events and their individual errors (only for shower
  events). Here $+$ corresponds to shower event and $\times$
  sign corresponds to track event with their corresponding event ID. 
We have also shown the positions and names of the blazars which are within the
  median angular error of the  IceCube events and have a TS value $>
  0$. The objects in blue color are
  FSRQs and in red color are BL Lac.
}
\label{equ_coordi}
\end{figure}

The ANTARES analysis takes into account both the temporal and energy
dependence of the flaring events whereas our analysis is independent
of the time. The observed IceCube events can be modelled by taking into
account two hypothesis: (1) the
events could be produced by atmospheric muons and the muon neutrinos (background), or (2) 
from an astrophysical source which also includes the background contribution.
A good test of compatibility is the ratio of these two hypothesis. 
We can take the ratio of 
the likelihood with the background of unique weight
($n_s=0$) and the maximized likelihood of the second
hypothesis with the corresponding $n_s$ values defined as $n_s=n^*_s$.
Now to evaluate each point source we use this Test Statistic (TS)
taking minus twice the log of the likelihood
ratio, 
 \be
  TS = - 2\, \text{log} \left [ \frac{{\cal L} (n_s = 0)}{{\cal L} (n_s=n^*_s)}
    \right ].
 \ee
For this procedure we use a full-sky IceCube events. For our present
analysis, we take into account 36 events out of reported 37 events
(event 32 is excluded in the present analysis because its energy and direction are not reported). 
We calculate the significance of each source
location, running 10,000 simulations in which the declination of each
IceCube sample event is fixed but the right ascension is randomized.
The p-value is calculated as the number of simulations with
$TS_{(sim)} \geq TS$ divided by the total number of simulations for a
given source, where $TS_{(sim)}$ is the TS value
obtained from the simulation.  
Also, the posteriori p-value for each object is estimated as the
fraction of the randomized simulations that yields an equal or
higher TS value for at least one of the 41 ANTARES sources. The
compatibility of the second hypothesis depends on the estimate of the
posteriori p-value. If the posteriori p-value is close to unity then it is consistent with the background.

\begin{table*}
\centering
\begin{tabular}{lccccccccc}
\hline
Object & Type & ID &   {$\delta\chi^2$}&$E_{\nu}$/TeV & 
{$\epsilon_{\gamma}$/keV} & $n^*_s$ & TS & p &  post-p \\

{(RA,Dec.)} &   z, $\Gamma$  &     & &  &     &              &       &
value      &    value \\      
\hline
PKS2326-502\cite{mdutka-thesis2014}& FSRQ & 7 & 0.46 & 34.3 & 182.19 & 0.06 & 0.0008 & 0.44 & 1.0 \\
(352.32, -49.94) & 0.518,30  &\, &\, &\, &\, &\, &\, &\, &\, \\  
\hline
PKS0208-512\cite{Tavecchio:2002xa} & FSRQ & 7  & 0.29 & 34.3 & 37.67 & 0.59 & 0.109 & 0.22 &1.0 \\
(32.7, -51.2) & 1.003, 18 & \, & \, &\,&\,&\,&\,&\, &\,\\
\hline
PKS0235-618\cite{Krauss:2014tna} & FSRQ & 7,20 & 0.80,0.27 &34.3, 1141& 21.68, 0.65& 0.39 & 0.040 & 0.18 & 1.0 \\
(39.29, -61.62) & 0.467,10  & \, & \, &\,&\,&\,&\, &\, &\,\\
\hline
PMNJ2345-1555\cite{Ghisellini:2013naa}  & FSRQ &21 & 0.48 & 30.2 & 34.07-51.62 & 0.71 & 0.197 & 0.43 & 1.0 \\
(356.27, -15.89) & 0.621,13-16  & \, & \, &\,&\,& \, & \, & \, &\,\\ 
\hline
B22308+34 & FSRQ & 34 & 0.23 & 42.1 & - & 0.97 & 0.503 & 0.48 & 1.0 \\
(347.77, 34.43) & 1.817, - & \, & \, &\,&\,& \, & \, & \, &\,\\ 
\hline
PKS0244-470  & FSRQ & 7 & 0.54 & 34.3 & - & 0.73 &  0.179 & 0.17 & 1.0 \\
(41.06, -47.06) & 1.385, - & \, & \, &\,&\,& \, & \, & \, &\, \\ 
\hline
CTA102\cite{Hervet:2016sou}  & FSRQ & 34 & 0.31 & 42.1 & 8.85 & 0.77 & 0.249 & 0.53 & 1.0 \\
(338.12, 11.72) & 1.036, 10 & \, & \, &\,&\,& \, & \, & \, &\,\\ 
\hline
PMNJ2331-2148\cite{Ghisellini:2015eha} & FSRQ & 21 & 0.52 & 30.2 & 31.23 & 0.6 & 0.125 & 0.45 & 1.0 \\
(352.75 ,-21.74) & 0.563, 12  &\, &\, &\, &\, &\, &\, &\, &\,\\  
\hline
PKS2227-08\cite{Savolainen:2009ix} & FSRQ & 34 & 0.97 & 42.1 & 5.80 & 0.53 & 0.096 & 0.53 & 1.0 \\
(337.44, -8.55) & 1.559, 10  &\, &\, &\, &\, &\, &\, &\, &\, \\  
\hline
OJ287\cite{Tavecchio:2002xa} & BL Lac & 26  & 0.62 & 210 & 6.43 & 1.32 & 0.691 & 0.31 & 0.99 \\
(133.85,20.09) & 0.306 , 12 & \, & \, &\,&\,& 0.69&0.184&0.33&1.0\\
\hline
PKS0805-07 & BL Lac & 27  & 0.52 & 60.2 & 7.43 & 1.23 & 0.556 & 0.24 & 1.0 \\
(122.06, -7.85) & 1.837, 15 & \, & \, &\,&\,&0.54&0.102&0.27&1.0\\
\hline
3C454.3\cite{Hervet:2016sou} & FSRQ & 34 & 0.31 & 42.1 & 25.72 & 0.85 & 0.33 & 0.50 & 1.0 \\
(343.5, 16.15) & 0.859, 15 & \, & \, &\,&\,&0.24&0.022&0.49&1.0\\
\hline
\hline
\end{tabular} 
\caption{ 
The objects which are in the error circles of the
IceCube events (ID in third column) are given in the first column.
Below each object we also put their coordinates, Right Ascension and Declination ( R.A., Dec.) in
degrees (this table is given in equatorial coordinates). The second
column gives the type of object and below this we also give its
redshift (z) and the bulk Lorentz factor ($\Gamma$). In the 
fourth column, the $\delta\chi^2$ of the object is given. In the fifth
and the sixth columns the deposited neutrino energy $E_{\nu}/TeV$  and 
the corresponding seed photon energy $\epsilon_{\gamma}/keV$ are
given.  In columns seventh and eighth the values of the $n^*_s$ and TS
are given from the Maximum Likelihood Method. In columns ninth and tenth the p-value and the posteriori
p-value (post p-value) are also shown. The last three objects are without
(upper value)  and
with (lower value) the prompt contribution to the background PDF.
}
\label{tab1}
\end{table*}

\section{Results}

        In the context of recent IceCube results, we analysed the 41 flaring blazars
taken from the Fermi-LAT catalog which are previously studied by the ANTARES collaboration to
look for possible temporal and spatial correlation\cite{Adrian-Martinez:2015wis}. We have also
analysed the 7 TeV flaring objects as discussed by ANTARES
collaboration for
the possible spatial correlation with the IceCube events. In fact all
these 7 objects are there in the TeVCat\cite{TeVCat}  which we had already analysed
in Ref.\cite{Sahu:2014fua} and found that the only HBL, PG 1553+113 has the
positional correlation with the IceCube event 17. So we don't
discuss about these 7 flaring objects here any more. For our analysis
of the possible correlation of IceCube events with the ANTARES sources
we use the unbinned MLM and two different values of spectral index
$\kappa=2$ and 2.5. We also do the separate analysis with and without
the contribution from the prompt flux coming from the charm hadron
decay. Our results are summarised in Table \ref{tab1}.

All the 28 shower events with their individual
errors and the 8 track events are shown in the sky map with equatorial coordinates in
Fig. \ref{equ_coordi}. 
The positions of ten FSRQs and two BL Lac objects
are also shown in the sky map.

\subsection{ Spectral index $\kappa=2$}

From the  41 Fermi blazars of ANTARES list,  32 objects have TS $> 0$
for the spectral index $\kappa=2$ without the prompt contribution to
the background. However, this number reduces to 19 when we include the
charm contribution.

From the above 32 objects 12 are within the median angular error of at least one IceCube
event having $\delta\chi^2 \, < 1$. The FSRQ, PKS 0235-618 is the only object associated
with two IceCube events (7, 20).
The FSRQs, PKS 2326-502, PKS 0208-512, PKS 0235-618 and PKS
0244-470 are within the error circle of event 7, while the FSRQs, 
3C454.3, B22308+34, CTA102 and PKS 2227-08 are within the error circle
of event 34. Another two FSRQs, PMNJ2345-1555 and PMNJ 2331-2148 are
within the error circle of the IceCube event 21. The BL Lac objects,
OJ287 and PKS0805-07 are coincident with the events 26 and 27 respectively.
All the relevant parameters of the above objects are shown in  Table \ref{tab1}.

The posteriori
p-value of all the above 12 objects are $\ge 99\%$. This shows that
our result (without the prompt contribution to the atmospheric
background) is consistent with the background fluctuation.

By including the prompt contribution to the background we
found that 19 objects have TS $> 0$ of which only three objects two  BL
Lac objects (OJ287, PKS 0805-07) and one FSRQ (3C454.3)  are within the median angular error of three IceCube
events (26, 27, 34). These three objects are shown in the table.

We observed that the background photon energy
$\epsilon_{\gamma}$ for most of the events  are below $<\, 40$ keV which shows
that the photon density $n'_{\gamma,f}$ can be large in the inner
region of the jet. 
By assuming a conservative $1\%$ energy loss by the
UHE protons we get the photon density in the inner region
$n'_{\gamma,f}\sim 2\times 10^{10}\, cm^{-3}$ which has a
radius $R'_f\sim 10^{15}\, cm$. Estimate of $R'_f$ value depends on
the outer blob radius $R'_b$, while the later parameter is adjusted to
fit the spectral energy distribution (SED) in the leptonic model of
the objects. However, for most of the objects $R'_b > 10^{15}$ cm is
taken to fit the SED\cite{Sahu:2014fua}. So, here  we take $R'_f\sim
10^{15}\, cm$ for the estimation of $n'_{\gamma,f}$. The
simulation shows that the $0 <  {\text TS} \,< 1$ for all the objects.

The diffuse neutrino flux $F_{\nu}$ for all
these objects is $2.31\times10^{-9}\,$
$GeV\,cm^{-2}\,s^{-1}\,sr^{-1}$ because for $\kappa=2$, the integral in Eq. (\ref{fluxnu}) is
independent of the redshift,  so the $F_{\nu}$ is the same for all the objects.
On the other hand it varies slightly for $\kappa\neq 2$. 
The high posteriori p-value for all these objects shows that our result
is consistent with the background fluctuation.
We also repeated the simulation for $\kappa=2.5$. As $\kappa$ changes the flux reduces
and found that non of the 41 objects satisfy the condition
  TS $>0$.

\section{Discussion }

ANTARES collaboration looked for possible temporal and spatial
correlation of 41 flaring objects selected from the Fermi-LAT catalog. We
analysed the same objects for the possible spatial correlation with
the IceCube events. For our analysis, we take into account the energy dependence of
both the background and the signal constructed from the data of the 79
IceCube string configuration. We
consider two different values of the spectral index $2$ and $2.5$ and
also analyse our results with and without the prompt contribution to
the atmospheric neutrino flux. We observed that,
from the 41 flaring objects, for $\kappa=2$, the MLM gives twelve
objects (without prompt flux contribution) and three objects (with
prompt flux contribution)  within the error circle of some IceCube events.
For these objects we have also estimated the neutrino
flux. However, for all these possible
candidates, the TS value is very small which leads to very high
posterior p-values $\ge 99\%$ and is consistent with the background
fluctuation. It is possible that the high energy neutrino flux from
these objects are much below the IceCube limit or blazars may not have
powerful central engine to produce very high energy cosmic rays. So 
most of the events in IceCube can be from some other type of sources. 
We have to wait for more data to look for possible
correlation of FSRQs and BL Lac objects with the IceCube events.

We  thank S. Mohanty for many useful comments and discussions. The
work of S. S. is partially supported by DGAPA-UNAM (Mexico) Projects
No. IN110815.

\end{document}